\documentclass[a4paper]{article}

\usepackage{INTERSPEECH2019}
\usepackage{graphicx}
\usepackage{hyperref}
\usepackage{amssymb}
\usepackage{textcomp}

\DeclareMathOperator*{\argmax}{arg\,max}

\title{End-to-End Multi-Speaker Speech Recognition using Speaker Embeddings and Transfer Learning}
\name{Pavel Denisov, Ngoc Thang Vu}
\address{Institute for Natural Language Processing (IMS), University of Stuttgart}
\email{\{pavel.denisov, thang.vu\}@ims.uni-stuttgart.de}

\begin{document}
%\ninept
%
\maketitle
\begin{abstract}
This paper presents our latest investigation on end-to-end automatic speech recognition (ASR) for overlapped speech.
We propose to train an end-to-end system conditioned on speaker embeddings and further improved by transfer learning from clean speech.
This proposed framework does not require any parallel non-overlapped speech materials and is independent of the number of speakers.
Our experimental results on overlapped speech datasets show that joint conditioning on speaker embeddings and transfer learning significantly improves the ASR performance.
\end{abstract}
\section{Introduction}
\label{sec:intro}
Recently, deep learning technology has boosted automatic speech recognition (ASR) performance significantly ~\cite{hinton2012asr, dahl2012cd, msr2016asr, povey2016purely}.
Overlapped speech -- well known in a more general context as the cocktail party problem -- remains, however, to be a largely unsolved problem.
Its difficulty can be mainly explained by high similarity of acoustic characteristics of signals that need to be separated and absence of any other obvious clues that might guide a potential solution.

Speech recognition of overlapped speech is usually approached in two stages: first overlapped speech is split to separate recordings of each speaker, then speech recognition
is performed on separated recordings.
Previous works on speech separation problem include
computational auditory scene analysis \cite{wang2006computational},
non-negative matrix factorization \cite{schmidt2006single},
graphical modeling \cite{hershey2010super}
and spectral clustering \cite{bach2006learning}.
Similarly to speech recognition, speech separation
methods have also made major progress
with the help of deep learning.
The deep clustering method has been introduced
in \cite{hershey2016deep} and consequently
improved in \cite{isik2016single,wang2018alternative}
and has become one of the most remarkable
speech separation methods in the recent years.
Deep clustering operates on a spectrogram
of overlapped speech and learns to map
time-frequency (T-F) units to high dimensional
embedding space.
More recently, a different approach,
named VoiceFilter, has been proposed in \cite{wang2018voicefilter}.
The work simplifies the problem of multiclass classification
of T-F units over multiple speakers to binary classification
between target speaker's speech and everything else.
To condition the neural network on specific speaker, the input is extended with a speaker embedding vector,
which is extracted by a separately trained network from the reference clean speech.

A new type of ASR systems, called end-to-end ASR, has emerged in the past years \cite{hannun2014deep, miao2015eesen, bahdanau2016end, chan2016listen, watanabe2017hybrid}.
End-to-end ASR maps the acoustic signal to written language
with a single encoder-decoder recurrent neural network
and does not require any domain specific knowledge
for solving of intermediate subtasks, such as grapheme-to-phoneme conversion.
Recently, two works have proposed to integrate speaker separation stage to end-to-end ASR.
The first one \cite{settle2018end} connects pretrained deep clustering model and end-to-end
ASR for the subsequent join fine-tuning for
the better ASR results.
The second one \cite{seki2018purely} removes explicit
speech separation part and trains end-to-end ASR
for simultaneous speech separation and recognition
by permutation invariant procedure, in which ASR system is optimized for multiple outputs
corresponding to multiple speakers in the input mixture.
Joint speech separation and recognition is also described in \cite{chen2018progressive}, although this work is based on the conventional ASR.
Both \cite{seki2018purely} and \cite{chen2018progressive} suggest that transfer learning from clean speech ASR improves the results of overlapped speech ASR.

Our work blends the ideas from
\cite{zmolikova2017,wang2018deep,wang2018voicefilter,settle2018end,seki2018purely, chen2018progressive}
and proposes to train an end-to-end overlapped speech recognition
system conditioned by speaker embeddings and
improved by transfer learning from clean speech.
It has advantages over \cite{wang2018voicefilter} \cite{settle2018end} and \cite{chen2018progressive} of not requiring parallel clean speech material and over \cite{seki2018purely} of not depending on the number of speakers.
We evaluate our proposed framework on overlapped speech datasets with two and three overlapped speakers, within and across settings.
Overall, we observe significant improvements over the baseline end-to-end ASR system.

\section{Method}
\label{sec:method}
\begin{figure}[ht]
\centering
\includegraphics[width=0.24\textwidth]{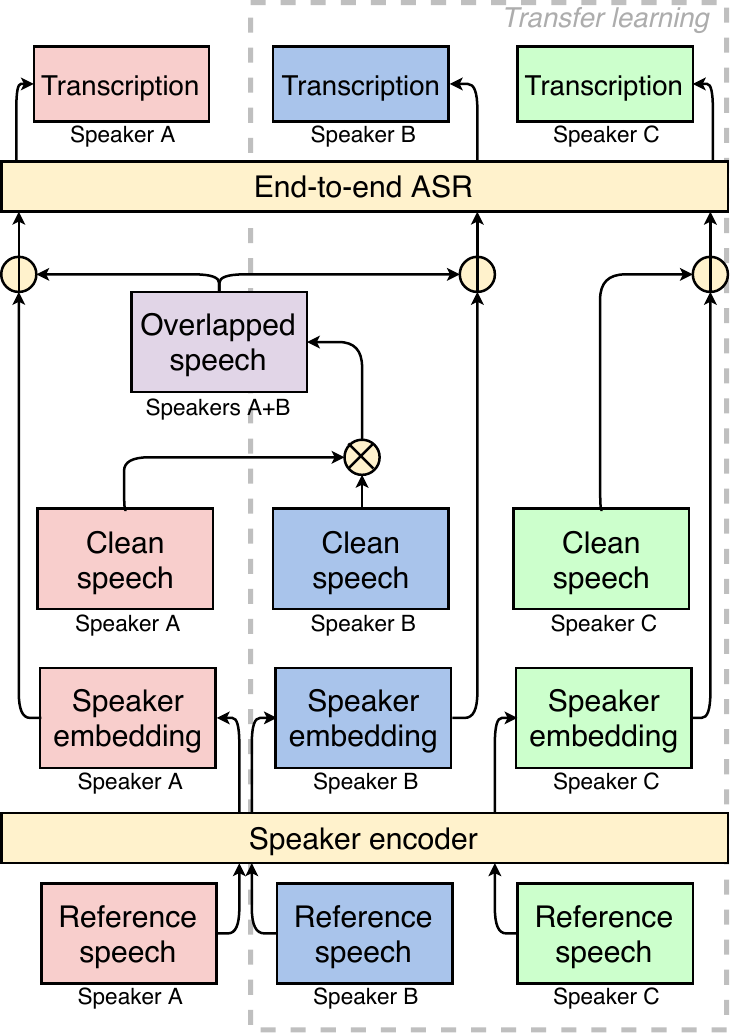}
\caption{
Overview of end-to-end ASR using speaker embeddings and transfer learning.
}
\label{fig:system}
\end{figure}

The outline of the proposed method is presented on Figure \ref{fig:system}.
It shows two separate neural network models, the speaker encoder and
the end-to-end ASR, together with their inputs and outputs.
The speaker encoder takes reference speech utterances on the input
and produces speaker embedding vectors, as described in sections \ref{ssec:embedding} and \ref{ssec:xvectors}.
The end-to-end ASR takes acoustic features of overlapped speech
with speaker embedding vector of the target speaker on the input and generates transcription
of the target speaker's speech on the output, which is used to update the parameters
of end-to-end ASR model during the training or provided as the final output
during the decoding.
A more detailed description of end-to-end ASR is given in sections \ref{ssec:e2e-asr} and \ref{ssec:baseline}.
Optionally, clean speech can be used
during the training in order to perform a basic form of transfer
learning, which is described in section \ref{ssec:transfer}.

\subsection{End-to-end ASR}
\label{ssec:e2e-asr}
End-to-end ASR has a hybrid CTC/attention architecture,
described in detail in \cite{watanabe2017hybrid}.
The input of the model is defined as a $T$-length sequence of
$d$ dimensional feature vectors $X = \{ x_t \in \mathbb{R}^d | t = 1, \dots, T \}$,
and the output of the model is defined as a $U$-length sequence
of output labels $Y = \{ y_u \in \mathcal{U} | u = 1, \dots, U \}$,
where $\mathcal{U}$ is a set of distinct output labels
and usually $U < T$.
During the training, a weighted sum of CTC loss and attention-based
cross-entropy loss is minimized:
\begin{align}
\mathcal{L} = \lambda \mathcal{L}_{\text{ctc}} + (1-\lambda) \mathcal{L}_{\text{att}},
\end{align}
where $0 \leq \lambda \leq 1$.
Attention-based cross-entropy loss is calculated according to predictions
of the attention-based encoder-decoder network:
\begin{align}
\mathcal{L}_{\text{att}} = & - \log \mathnormal{p}_{\text{att}}(Y|X) \\
\mathnormal{p}_{\text{att}}(Y|X)  = & \prod_u \mathnormal{p}(y_u | X, y_{1:u-1}) \\
\mathnormal{p}(y_u | X, y_{1:u-1}) = & \text{ Decoder} (\mathbf{r}_{u}, \mathbf{q}_{u-1}, y_{u-1}) \\
\mathbf{h}_t = & \text{ Encoder}(X) \\
a_{ut} = & \text{ Attention}(\{a_{u-1}\}_t, \mathbf{q}_{u-1}, \mathbf{h}_t) \\
\mathbf{r}_u = & \sum_t a_{ut} \mathbf{h}_{t}.
\end{align}
Here, $\text{Encoder}(\cdot)$ and $\text{Decoder}(\cdot)$ are
recurrent neural  networks, $\text{Attention}(\cdot)$
is an attention mechanism and
$\mathbf{h}_t$, $\mathbf{q}_{u-1}$ and $\mathbf{r}_u$
are the hidden vectors.
CTC loss is calculated from a linear transformation of encoder output
and all possible $T$-length sequences of an extended output labels set
$Z = \{ z_t \in \mathcal{U} \cup \texttt{<blank>} | t = 1, \dots, T \}$,
corresponding to sequence $Y$ of original output labels:
\begin{align}
\mathcal{L}_{\text{ctc}} & = - \log \mathnormal{p}_{\text{ctc}}(Y|X) \\
\mathnormal{p}(Y|X) & \approx \underbrace{\sum _{Z} \prod _t \mathnormal{p}(z_t | z _{t-1}, Y) \mathnormal{p}(z_t|X)} _{\triangleq \mathnormal{p} _{\text{ctc}}(Y|X)} \mathnormal{p}(Y) \\
\mathnormal{p}(z_t | X) & = \text{ Softmax}(\text{Lin}(\mathbf{h}_t) ).
\end{align}
During the decoding, the same CTC and
attention-based probabilities are summed with possibly
another weight and are used to find
the most probable output labels sequence:
\begin{align}
\hat{Y} = \argmax_Y \{ \lambda \log \mathnormal{p}_{\text{ctc}}(Y|X) + (1-\lambda) \log \mathnormal{p}_{\text{att}}(Y|X) \}.
\end{align}

\subsection{Speaker embeddings}
\label{ssec:embedding}
Speaker embedding is a vector of fixed dimensionality that
represents speaker's characteristics and can be extracted
from a reference recording of speaker's speech.
Speaker embeddings have been shown to be a useful source of information
about speaker in many tasks, including speaker verification, speaker diarization \cite{wang2018speaker},
speech synthesis \cite{jia2018transfer}
and speech separation \cite{wang2018voicefilter}.
We condition ASR system for the recognition
of speech of a certain speaker in the recording
of overlapped speech.
This arrangement removes a major roadblock
of the permutation problem,
appearing when multiple correct outputs are possible
for a single input,
and simplifies application of a wide range of well studied
machine learning methods for further system optimization.

\subsection{Transfer learning}
\label{ssec:transfer}
Speech recognition of overlapped speech can be viewed
as speech recognition of clean speech in mismatched
conditions and be addressed by numerous transfer learning methods.
It has been proposed in \cite{chen2018progressive} to utilize
teacher-student training for transfer learning
from clean to overlapped speech recognition.
This approach has a limitation of applicability
on training sets with parallel clean and overlapped speech only,
which is hardly achievable in real life scenarios.
Another example of transfer learning from clean
to overlapped speech recognition is given in \cite{seki2018purely}, which applies parameters transfer.
Multi-condition training is an alternative to the parameters transfer method,
in which training samples from different conditions
are mixed together and used for the training simultaneously.
Multi-condition training has been demonstrated to provide better results for transfer learning between languages
\cite{heigold2013multilingual} as well as between channels \cite{ghahremani2017investigation}.
In this work, we experiment with both parameters transfer and
multi-condition training methods
in order to improve overlapped speech recognition by
utilization of clean speech training data.

\section{Experimental setup}
\label{sec:experimental}

\subsection{Datasets}
\label{ssec:datasets}
We evaluate our models on the widely used mixed speech datasets
wsj0-2mix and wsj0-3mix \cite{hershey2016deep, isik2016single}.
The datasets contain two-speaker and three-speaker mixtures
of randomly selected utterances from WSJ0 corpus.
Training, development and evaluation sets,
named \texttt{tr}, \texttt{cv} and \texttt{tt},
are generated from WSJ0 training, development and evaluation sets
\texttt{si\_tr\_s}, \texttt{si\_dt\_05} and \texttt{si\_et\_05}
and contain speech of speakers of
both genders in different combinations.
We use \texttt{max} version of the datasets, meaning that
the length of every mixed speech utterance
is chosen to be maximum of the lengths of original
utterances being used for a mixture.
The sampling rate of the used datasets is 16 kHz.
Training, development and evaluation sets
of both datasets contain 20000, 5000 and 3000 utterances.
Training and development sets contain speech of same 101 speakers,
therefore development sets are used for the evaluation
in closed speaker set condition.
Evaluation sets contain speech of another 19 speakers,
therefore evaluation sets are used for the evaluation
in open speaker set condition.
Total durations of training, development and evaluation sets of
wsj0-2mix dataset are 46, 11 and 7 hours.
Total durations of training, development and evaluation sets of
wsj0-3mix dataset are 51, 13 and 8 hours.
LibriSpeech \cite{panayotov2015librispeech}
\texttt{train-clean-100} dataset is used for the
transfer learning experiments. It contains 28539 utterances
of read speech by 251 speakers and has total duration of 100 hours.
The sampling rate of the dataset is 16 kHz.

\subsection{Baseline}
\label{ssec:baseline}
Our end-to-end ASR system is based on
ESPnet toolkit \cite{watanabe2018espnet} and its WSJ recipe.
80-dimensional log Mel filterbank coefficients with
pitch with a frame length of 25 ms and shift of 10 ms are used as input features.
The input features are extracted and normalized
to zero mean and unit variance
with Kaldi toolkit \cite{povey2011kaldi}.
The encoder network consists of four BLSTM layers with 320 units
in each layer and direction and linear projection layer with 320 units.
No subsampling is applied to the input.
The decoder network consists of one LSTM layer with 300 units.
Additive attention mechanism with 320 dimensions is utilized.
We use 49 characters as output units.
PyTorch backend 
of ESPnet is used to implement the networks.
Training is performed with
AdaDelta optimizer \cite{zeiler2012adadelta}
and gradient clipping 
on two GPUs in parallel with a batch size of 30 for 30 epochs.
The optimizer is initialized with $\rho = 0.95$ and $\epsilon = 10^{-8}$.
$\epsilon$ is halved after an epoch if performance of the model
did not improve on development set.
The model with the highest accuracy on development set
is used for the decoding. The CTC weight $\lambda$
is set to $0.2$ during the training and $0.3$ during the decoding.
The decoding is performed with a beam search
with a beam size of 30.

The decoding makes use of word-based
RNN-LM \cite{hori2018end} with a weight of $1.0$.
Word-based RNN-LM is trained on the LM training set of WSJ0 corpus
containing 37M words and 1.6M sentences, dictionary size is 65K words.
Word-based RNN-LM contains 1 LSTM layer with 1000 units.
The stochastic gradient descent optimizer is used to train word-based
RNN-LM with a batch size of 300 for 20 epochs.

\subsection{Speaker embeddings extractor}
\label{ssec:xvectors}
We extract 512-dimensional speaker embeddings
from the reference utterances with x-vector system from Kaldi toolkit.
We use the pretrained model downloaded from \url{http://kaldi-asr.org/models/m8}.
The model is trained on augmented
VoxCeleb 1 \cite{nagrani2017voxceleb} and VoxCeleb 2 \cite{chung2018voxceleb2}
datasets according to the procedure
closely following the description in \cite{snyder2018x}
and evaluated on Speakers in the Wild dataset \cite{mclaren2016speakers}
with 3.5\% equal error rate. The input features of x-vector extractor
are 30-dimensional MFCCs without cepstral truncation
with a frame length of 25 ms and shift of 10 ms.
Mean normalization with a sliding window
of up to three seconds is applied to the input features.
Speaker embeddings are extracted from voiced frames only,
which are selected by the same energy-based VAD.
We obtain one vector from each utterance
and average them per speaker to get speaker specific vector.
$L_2$ normalized speaker embedding
is used as an additional input for the ASR model.
We denote insertion of the speaker embedding vector to the beginning
of the sequence of acoustic features vectors as \textit{horizontal stacking}.
We denote concatenation of the speaker embedding vector with every
acoustic features vector as \textit{vertical stacking}.
Horizontal stacking requires speaker embedding and acoustic features to have same size,
which can be achieved either by downscaling
of the speaker embedding vector, which we perform with
the trainable linear transformation,
or by padding of the acoustic feature vectors,
which we perform by appending appropriate number of zeros
to each acoustic features vector.
Vertical stacking allows the size of the speaker embedding
to be independent of the size of the acoustic features.

\section{Results}
\label{sec:results}
\subsection{Speaker embeddings inclusion strategies}
The first set of experiments aims to determine the best
strategy for inclusion of speaker embeddings in the model's input.
While vertical stacking does not enforce same size of the speaker embedding
and acoustic features, we perform
two experiments with vertical stacking: the first one with unchanged sizes of the input vectors
and the second one with the downscaled speaker embedding.
The second experiment isolates
the effect of different stacking types
from the effect of the speaker embedding downscaling.
50 utterances of reference speech are used to produce each speaker embedding in this experiment.
It can be seen from the data in Table \ref{tab:stacking} that
vertical stacking clearly outperforms horizontal stacking,
while the downscaling of the speaker embedding
results in minor degradation of the model's performance.

\begin{table}[h]
  \caption{ Results (WER, \%) with different speaker embedding inclusion strategies on the two-speaker overlapped speech dataset. }
  \label{tab:stacking}
  \centering
  \footnotesize
  \begin{tabular*}{1.0\columnwidth}{l|c|c}
    \noalign{\hrule height 1pt}
    Strategy & dev & eval \\
    \hline
    Baseline (no speaker embeddings) & 79.6 & 85.7 \\
    Horizontal stacking with downscaled embedding & 77.0  & 84.1 \\
    Horizontal stacking with padded acoustic features & 83.0  & 88.6 \\
    Vertical stacking with downscaled embedding & 11.7  & 24.9 \\
    Vertical stacking with unchanged vectors' sizes & \textbf{11.4} & \textbf{22.1} \\
    \noalign{\hrule height 1pt}
  \end{tabular*}
\end{table}

\subsection{Amount of reference speech data}
The next set of experiments is concerned with amount of reference speech data
required for the generation of speaker embeddings.
Table \ref{tab:ref} gives an overview of ASR performance with different numbers of reference utterances
in case of two-speaker overlapped speech.
It is apparent from this table that larger amount of reference speech data
allows to generate more general speaker embeddings
preventing the ASR model from the overfitting towards known speakers
and seen utterances.
It is worth noting, however, that even one reference utterance of the duration of approximately ten 
seconds appears to be sufficient material for the major improvement.

\begin{table}[h]
  \caption{ Results (WER, \%) with different amount of reference speech data on the two-speaker overlapped speech dataset. }
  \label{tab:ref}
  \centering
  \footnotesize
  \begin{tabular*}{0.9\columnwidth}{l|l|l|c|c}
    \noalign{\hrule height 1pt}
    \multicolumn{3}{c|}{Reference speech amount per speaker} & dev & eval \\
    \cline{1-3}
    Utterances & Seconds & Voiced frames &  & \\
    \hline
    1  & 8.3 \textpm 2.9     & 641 \textpm 271  & 15.8 & 32.6  \\
    5  & 42.3 \textpm 6.4    & 3156 \textpm 623  & 17.3 & 29.1  \\
    10 & 84.6 \textpm 10.1   & 6341 \textpm 1090 & 11.3 & 22.6 \\
    20 & 170.3 \textpm 16.9  & 12545 \textpm 1783  & \textbf{10.8} & 22.5 \\
    50 & 426.3 \textpm 34.2  & 31490 \textpm 4260  & 11.4 & \textbf{22.1} \\
    \noalign{\hrule height 1pt}
  \end{tabular*}
\end{table}

\subsection{Transfer learning}
We evaluate two transfer learning approaches,
namely parameters transfer and multi-condition training,
for improving ASR performance on overlapped speech
by utilization of non-parallel clean speech training data.
Table \ref{tab:mix} presents the results of the systems trained on
training sets of wsj0-2mix and wsj-3mix datasets
and tested on development and evaluation sets
of the corresponding datasets.
The most striking observation to emerge from the results is
that the training process has not converged
on the dataset with three overlapping speakers,
but the system has been able to decode the same recordings
when trained on the combination of overlapped and clean speech datasets.
This finding can be attributed to higher
complexity of the modeled function in case of
increased number of overlapping speakers,
which the neural network could not learn
just from the overlapped speech data,
and demonstrates how crucial the role of
the transfer learning approach
for the solution of certain problems can be.
The transfer learning results are expectedly better on
the dataset with two overlapping speakers as well,
especially for the open speaker condition,
what is due to the number of additional speaker embeddings
in the training data and subsequent better generalization of
the relationship between speaker embedding and relevant acoustic features.
Finally, multi-condition training has demonstrated slightly
better results than parameters transfer.
This finding is in agreement with the previous
reports on transfer learning applications in ASR.

\begin{table}[h]
  \caption{ Results (WER, \%) of the baseline ASR, speaker embeddings conditioning and transfer learning on the two- and three-speaker overlapped speech datasets. }
  \label{tab:mix}
  \centering
  \footnotesize
  \begin{tabular*}{0.91\columnwidth}{l|c|c|c|c}
    \noalign{\hrule height 1pt}
    & \multicolumn{2}{|c|}{wsj0-2mix} & \multicolumn{2}{|c}{wsj0-3mix} \\
    \cline{2-5}
    & dev & eval & dev & eval \\
    \hline
    Baseline & 79.6 & 85.7 & 95.9 & 96.0 \\
    + speaker embeddings & 11.4 & 22.1 & 95.6 & 95.7 \\
    \hspace{2mm} + parameters transfer & 8.8 & 16.9 & 22.7 & 45.3 \\
    \hspace{2mm} + multi-condition training  & \textbf{8.5} & \textbf{14.6} & \textbf{21.7} & \textbf{42.9} \\
    \noalign{\hrule height 1pt}
  \end{tabular*}
\end{table}

As our method does not utilize any explicit knowledge about the number of
overlapping speakers, the models should theoretically also work
for testing data with larger or smaller number of overlapping speakers
than in the training data. We test whether this is true in practice
by decoding  wsj0-3mix testing data with the model trained
on wsj0-2mix training data (combined with LibriSpeech 100)
and vice versa. Encouraged by the success of the previous transfer learning experiments,
we also train a system on a combination of clean and
overlapped two- and three-speaker datasets.
The results of testing with
mismatching number of overlapping speakers are given in Table \ref{tab:speakers}.
In general it seems that the proposed method
does not depend on the number of overlapping speakers
but can benefit from training on larger amount of
different conditions of speech overlap.
A possible explanation of the slightly worse result of the best system on
the open speaker set condition with two overlapping speakers
might be a bias towards WSJ0 speakers
in the combined training dataset introduced by the addition
of wsj0-3mix dataset, which is slightly
larger than wsj0-2mix dataset and therefore has more impact
on two-speaker testing data than wsj0-2mix on three-speaker testing data
in this experiment.

\begin{table}[h]
  \caption{ Results (WER, \%) of the proposed method depending on the training data. }
  \label{tab:speakers}
  \centering
  \footnotesize
  \begin{tabular*}{0.98\columnwidth}{l|c|c|c|c}
    \noalign{\hrule height 1pt}
    Training data & \multicolumn{2}{|c|}{wsj0-2mix} & \multicolumn{2}{|c}{wsj0-3mix} \\
    \cline{2-5}
     & dev & eval & dev & eval \\
    \hline
    LibriSpeech 100 + wsj0-2mix & 8.5 & \textbf{14.6}  & 45.2 & 55.3 \\
    LibriSpeech 100 + wsj0-3mix & 7.8 & 28.5  & 21.7 & 42.9 \\
    LibriSpeech 100 + wsj0-\{2,3\}mix & \textbf{4.8} & 15.2 & \textbf{15.5} & \textbf{32.3} \\
    \noalign{\hrule height 1pt}
  \end{tabular*}
\end{table}

\subsection{Comparison with earlier work}

Table \ref{tab:previous} compares our best result on the evaluation set of wsj0-2mix dataset
with the results reported on the same dataset in the previous works.
From the table we can see that the proposed system outperforms the best known result by 42\% relatively.
However, it should be noted that the listed systems differ from each other
in a number of ways, including types of ASR system, types of LM
and types and amount of training data.

\begin{table}[h]
  \caption{ Results (WER, \%) of the proposed method and previous works on the two-speaker overlapped speech dataset. }
  \label{tab:previous}
  \centering
  \footnotesize
  \begin{tabular*}{0.9\columnwidth}{l|c}
    \noalign{\hrule height 1pt}
    Method & eval \\
    \hline
    Deep Clustering, hybrid ASR \cite{isik2016single} & 30.8 \\
    Permutation Invariant Training, hybrid ASR \cite{qian2018single} & 28.2 \\
    Permutation Invariant Training, end-to-end ASR \cite{seki2018purely} & 28.2 \\
    Speaker Parallel Attention, end-to-end ASR \cite{chang2018end} & 25.4 \\
    Proposed, end-to-end ASR & \textbf{14.6} \\
    \noalign{\hrule height 1pt}
  \end{tabular*}
\end{table}

\begin{figure}[htb]
\begin{minipage}[b]{1.0\linewidth}
  \centering
  \centerline{\includegraphics[width=8.5cm]{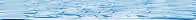}}
  \centerline{(a) Without speaker embedding}\medskip
\end{minipage}
\begin{minipage}[b]{1.0\linewidth}
  \centering
  \centerline{\includegraphics[width=8.5cm]{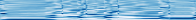}}
  \centerline{(b) With the speaker embeddings for the speaker \texttt{01z}}\medskip
\end{minipage}
\begin{minipage}[b]{1.0\linewidth}
  \centering
  \centerline{\includegraphics[width=8.5cm]{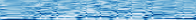}}
  \centerline{(c) With the speaker embeddings for the speaker \texttt{20h}}\medskip
\end{minipage}
\caption{
Visualization of the hidden vector sequences
for the utterance \texttt{01zc020o\_2.3474\_20hc010j\_-2.3474} of wsj-2mix dataset.
}
\label{fig:hidden}
\end{figure}
Following \cite{seki2018purely}, we also visualize the encoder networks outputs of an example utterance (see Figure \ref{fig:hidden}).
We apply principal component analysis to the hidden vectors
on the vertical axis as well.
Figure \ref{fig:hidden}(a) shows the output of the baseline's encoder,
while figures \ref{fig:hidden}(b) and \ref{fig:hidden}(c)
show the encoder network conditioned on speaker embeddings
of two different speakers. Some patterns
from the encoder's output of the baseline model appear on the conditioned
encoder's output for the first speaker, and another ones
appear on the encoder's output for the second speaker.
This observation suggests that the conditioning on the speaker embeddings
indeed allows the encoder network to perform the separation of
overlapped speech.

\section{Conclusions}
\label{sec:conclusions}
In this paper, we proposed an effective end-to-end speech recognition framework for overlapped speech using speaker embeddings and transfer learning techniques.
Experimental results on simulated overlapped speech datasets revealed that using speaker embeddings our framework was able to automatically identify relevant information of the target speaker for recognition.
The application of transfer learning technique played a crucial role with the increasing number of speakers.
Finally, we observed significant improvements over the baseline end-to-end system even while using just ten seconds of reference speech per speaker.

\bibliographystyle{IEEEtran}
\bibliography{refs}

\end{document}